# Bio-Inspired 4D-Printed Mechanisms with Programmable Morphology

Anurag Bhattacharyya[1], Jin-Young Kim[2], Lee R. Alacoque[3] and Kai A. James[4]*

[1]Palo Alto Research Center; Palo Alto, California, USA.

[2]Seoul National University; Seoul, South Korea.

[3]University of Illinois Urbana-Champaign; Urbana, Illinois, USA.

[4]Georgia Institute of Technology, Atlanta, Georgia, USA

*Corresponding author. Email: kai.james@gatech.edu

**Abstract**

Traditional robotic mechanisms contain a series of rigid links connected by rotational joints that provide powered motion, all of which is controlled by a central processor. By contrast, analogous mechanisms found in nature, such as octopus tentacles, contain sensors, actuators, and even neurons distributed throughout the appendage, thereby allowing for motion with superior complexity, fluidity, and reaction time. Smart materials provide a means with which we can mimic these features artificially. These specialized materials undergo shape change in response to changes in their environment. Previous studies have developed material-based actuators that could produce targeted shape changes. Here we extend this capability by introducing a novel computational and experimental method for design and synthesis of material-based morphing mechanisms capable of achieving complex pre-programmed motion. By combining active and passive materials, the algorithm can encode the desired movement into the material distribution of the mechanism. We demonstrate this new capability by de novo design of a 3D printed self-tying knot. This method advances a new paradigm in mechanism design that could enable a new generation of material-driven machines that are lightweight, adaptable, robust to damage, and easily manufacturable by 3D printing.

**Introduction**

Robotic mechanisms are among the most challenging systems to engineer due to their highly complex, multidisciplinary, and dynamic nature. However, biological organisms have mastered this task using a variety of approaches. This is one of many domains in which nature serves as a source of inspiration and a benchmark against which we measure our progress. Historically, engineered robots have contained rigid links connected via complex actuators such as electric motors [1]. This combination of rigidity and component complexity makes these systems susceptible to failure, and also makes them less maneuverable, particularly in tight spaces. By contrast, analogous mechanisms found in nature, such as an elephant trunk or an octopus tentacle, exhibit fluid motion and can be contorted to attain a much wider variety of possible configurations (See Fig. 1). Furthermore, whereas traditional robotics prioritizes the position and orientation of the hand or *end-effector* [2], there are many scenarios in which the full path of the arm may be critical. An example of such a scenario is shown in Fig. 1c, which features an octopus using its contorting ability to escape a jar [3].

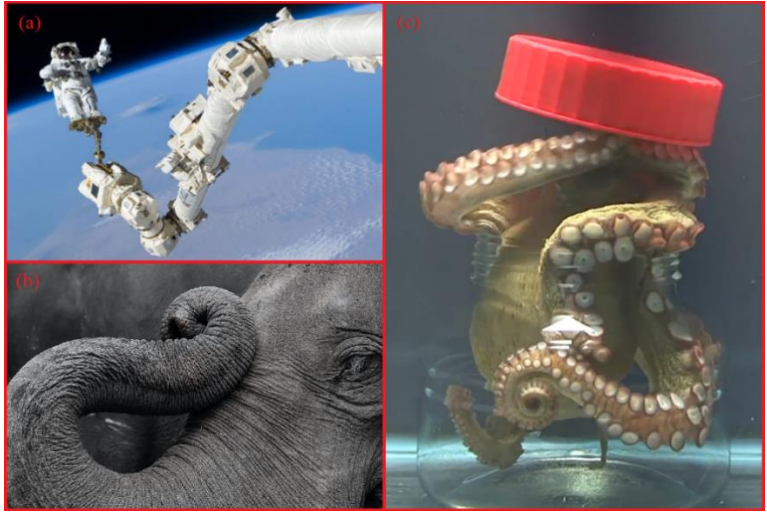

**Figure 1**: (a) The Canadarm robotic arm positioning an astronaut in space (photo courtesy of NASA); (b) An elephant contorting its trunk *[5]*; (c) An octopus using its tentacles to unscrew and escape from a jar *[3]*.

By replacing traditional rigid links and actuators with *smart materials*, we can ultimately create mechanisms whose robustness and maneuverability more closely resemble that of natural organisms. Smart materials exhibit shape changes in response to changes in their environment, such as temperature change. By exploiting this capability, our study seeks to advance a new paradigm in mechanism design. The resulting mechanisms will have no electromechanical components and no central processing station. Therefore, they can be rapidly and cheaply manufactured via 3D printing. This form of 3D printing in which the fabricated object is designed to change shape over time is sometimes referred to as *4D printing* [4]. Additionally, because the motion is no longer restricted to a small number of discrete actuators, 4D printed mechanisms can have a high concentration of degrees of freedom, similar to elephant trunks, which can have 40,000 muscles with no bones or joints, thus enabling highly complex motion [5]. This allows for increased maneuverability and much more complex motion.

Recently the concept of physical artificial intelligence has been introduced as counterpart to digital artificial intelligence [6]. Physical AI systems rely heavily on the use of smart composite materials to create next-generation robots that are akin to biological organisms. While researchers have made great strides in digital AI for robotics over the past few decades, advances in the development of robots' bodies, materials and morphology have not kept pace [6]. However, recent advances in 4D printing and computational design suggest that this emerging research area will have a significant role to play in filling this void.

Several early studies on 4D printing combined multiple smart materials, typically shape-memory polymers, to create hinge mechanisms that provided in-plane bending actuation. Ge *et al.* employed this strategy to create 4D printed active origami composites [7]. They were able to combine multiple hinge components to create mechanisms such as a self-assembling box and a mock paper airplane, both of which emerged from the 3D printer as a flat sheet before actively morphing into their respective 3D shapes. Raviv *et al.* used a similar approach in which they systematically assembled 4D printed actuators, which they referred to as primitives, to create long chains that could change shape and spell out pre-programmed acronyms [8]. Here, the researchers used an algorithm to determine the sequence of primitives necessary to obtain the desired end shape. More recently, Gu *et al*. used design optimization to create a soft prosthetic hand containing fibre-reinforced elastomeric composites [9]. Unlike the 4D printed examples cited above, motion at the hand's joints was achieved via pneumatic actuation. By replacing the electric motors found in conventional prosthetics, the prosthetic hand achieved significant cost and weight savings. Each of these examples demonstrates the power of material-driven actuation, but they maintain key features of traditional robotic mechanisms, namely the use of rigid struts separated by a small number of deformable joints. Consequently the number of achievable configurations for each designed mechanism remains limited. A more recent study by Gladman *et al*. took a biomimetic approach to create 4D printed plant-inspired designs that morph into target shapes encoded into their material architectures using composite hydrogels with controlled anisotropic swelling [10].

Topology optimization provides a computational framework for design of smart mechanisms with complex programmable motion. This technique was originally developed in the late 80's as a means of generating maximally stiff load-bearing structures [11] . Since its introduction, the method has been adapted to create a wide range of mechanical systems including compliant mechanisms [12], aeroelastic structures [13], cellular materials [14], and active composites [15]. Topology optimized smart composites contain a combination of active and passive materials distributed freely throughout the volume of the mechanism [16], such that there is no distinction between joints and structural members. Hence the entire body of the mechanism acts as an actuator, thereby allowing for complex free-form motion, similar to that which we see in nature. Several studies have proposed the use of topology optimization to automate the process of distributing active and passive material phases at the scale of the voxel, with 3D printing being used to fabricate the highly intricate designs that result [15] [17] [18]. In this way, the movement of the mechanism is programmed into the material distribution. However, the computationally intense nature of the topology optimization method has limited the complexity and programmability of the active mechanisms designed in these earlier studies. In each of the previously published examples, the motion of the mechanism was characterized by small shape changes [17], or rotation about a single axis (i.e. pure bending [17] [16] or pure torsion [15] [16]).

Here we introduce a novel method that incorporates topology optimization into a broader hierarchical framework in which the algorithm selects the material layout at two scales: at the smaller scale the material is selected voxel by voxel, while at the larger scale, the algorithm selects the arrangement of a series of optimized kinematic units. This approach allows us to generate large, complex motion while keeping the computational cost tractable. We extend the capability achieved in earlier studies by enabling complex motion characterized by large nonlinear deflections that include rotations about multiple reference axes. Thus, our designs combine the robustness and manufacturability of earlier 3D printed material-driven mechanisms with the programmability of conventional robotic mechanisms. Additionally, with the entire body of the mechanism providing sensing, actuation, and structural rigidity, this design method represents an important step toward creating engineered mechanisms that can compete with natural organisms in terms of adaptability and free-form morphability.

**Methods**

The proposed design method proceeds in three stages: 1) topology optimization of basic kinematic elements, 2) experimental characterization, and 3) computational assembly of the mechanism (see Fig. 2). In the first stage, we use multimaterial topology optimization [19] to design a series of kinematic elements that produce fundamental displacement outputs such as bending and twisting. The overall mechanism is an assembly of unitary elements that comprise three distinct classes: bending elements, torsional (i.e. twisting) elements, and *neutral* elements, whose shape remains fixed during the activation of the mechanism. Together, these elements enable rotational displacement about all three coordinate axes in both the positive and negative directions. Prior to actuation, each element is in the shape of a rectangular prism, and the optimization algorithm is used to distribute two smart materials throughout this prescribed volume in order to maximize the element's angular displacement. By dividing the design task into two stages (i.e. topology optimization and mechanism assembly), we significantly reduce the computational cost of the problem. Performing time-dependent three-dimensional finite element analysis (FEA) and optimization of the full mechanism in a single step would be highly computationally expensive. Our approach reduces the number of degrees of freedom in the FEA model by more than an order of magnitude.

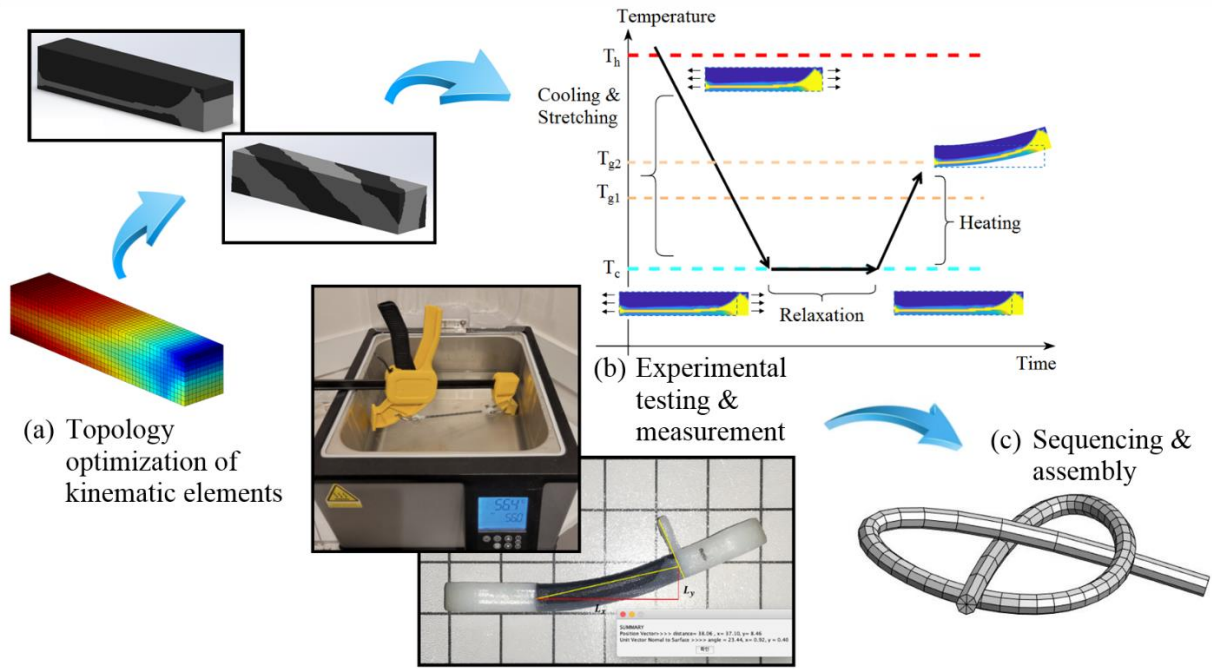

**Figure 2:** The hierarchical design framework (a) Structural topology optimization of kinematic elements; (b) The thermomechanical programming cycle used to trigger the shape-memory effect in the polymeric design materials. The specimen is initially heated to a hot temperature $T_h$, which is above the glass transition temperatures of both design materials ($T_{g1}$ and $T_{g2}$). It is then stretched and cooled to a temperature $T_c$, which is below both glass transition temperatures. It then undergoes a relaxation period during which the stretching force is removed. Lastly, it is reheated to a temperature between $T_{g1}$ and $T_{g2}$ to trigger the desired motion. Note that the blue and yellow images represent the topology optimized bending beam, with the dashed blue outline indicating the original shape and size prior to thermomechanical loading. The displacement is then measured via digital image correlation; (c) In the final stage of the process, we perform a second optimization procedure to obtain the required sequence of elements before fabricating the full mechanism.

The mechanism is composed of two shape-memory polymer materials. Although both materials contain shape-memory properties, one of the design materials is chosen to have a higher transition temperature so that when the mechanism is heated to activate the shape-memory effect, this higher temperature material remains unactivated. The resulting difference in the displacement response within the activated and unactivated material regions is harnessed to produce intricate three-dimensional motion. Therefore, in this context, we refer to the material with the lower transition temperature as the *active* material, and we refer to material with the higher transition temperature as the *passive* material. Within the optimization algorithm, we use *finite element analysis* to simulate and predict the transient displacement response of each kinematic element. This information is then passed to an optimization algorithm to update the material distribution within the mechanism. This process is carried out iteratively until it converges to an optimal design. In this way, the algorithm systematically determines which material (active or passive) should be used to populate each voxel within the overall volume. The output of the algorithm is a voxel-by-voxel description of the desired material distribution. A detailed description of the topology optimization procedure is provided later in this section.

In the second stage of the process, we experimentally measure the displacement response of each kinematic element. For this task, we 3D print a representative prototype of each element class and measure the angular displacement of the element in response to the designed thermomechanical cycle. Although the optimization algorithm uses numerical simulation, which offers an approximate prediction of the displacement response, this simulation can be subject to modeling errors. By performing this measurement step, we can accurately characterize the anticipated displacement response of each kinematic element. In this way, we effectively negate any errors associated with the computational model, since only the experimental measurements are passed to the final stage of the process, where a second algorithm assembles the elements into a specific sequence.

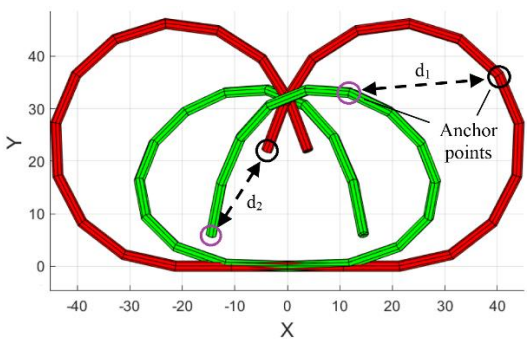

**Figure 3:** The designed self-tying knot (red) superposed onto the ideal knot (green). To find the optimal sequence of kinematic elements, we select two anchor points distributed along the length of the knot. Each point has an analogous point on the ideal knot located at a prescribed arc length from the root of the knot. The algorithm searches for the element sequence that causes the postion and orientation of the designed knot to be as close as possible to the position and orientation of the ideal knot at the designated anchor points.

To trigger the displacement response, each element must undergo a three-step thermomechanical programming cycle. In the final step of the cycle, we heat the element to an elevated temperature to activate its shape-memory effect and produce a self-morphing motion. The resulting displacement is then measured using a digital image correlation procedure. For this measurement, we imported the image file into the MATLAB software environment. From here, we manually selected key locations on the image, and we used a MATLAB script to extract the *xy*-coordinates of these points and compute the relevant displacement angles. Note that although we use a liquid bath for heating and activating the mechanism, this approach may not be practical for certain applications. The general methodology presented here is also compatible with other forms of heating, such as joule heating, which has been successfully applied to 3D printed polymers that have undergone a carbonization process to increase their electrical conductivity [20].

In the final stage of the design process, we use a genetic algorithm to determine the sequence of kinematic elements required to achieve the target displacement. In this study, we have chosen to design a self-tying knot as a demonstration example. This problem is particularly challenging because the knot path combines bending and torsion about multiple reference axes. The knot also exhibits large geometrically nonlinear displacements, which require a fully three-dimensional kinematic model, similar to that which is used in robotics problems [21]. We must also implement design constraints to prevent the various sections of the knot from colliding with one another during the motion. The genetic algorithm starts out by generating random sequences of kinematic elements and calculating the displacement trajectory of each sequence using a forward kinematics model. Each sequence serves as a candidate design for the self-tying knot mechanism. The calculated final shape of the sequence is then compared with the shape of the *ideal* knot, to quantify the *fitness* of each design (See Fig. 3). After successive iterations, the algorithm converges to produce a sequence whose end shape is as close as possible to that of the ideal knot. Below we provide a detailed description of the computational and experimental procedures implemented during this study.

*Topology Optimization*

To obtain the optimal material distribution within the kinematic elements, we perform a multimaterial topology optimization. Within the optimization algorithm, the thermomechanical response of the material is simulated using the finite element method (FEA). The FEA model for the 3D torsional unit contains a mesh of 625 8-node hexahedral elements. The FEA model for the bending element used a 2D mesh containing 2880 4-node quadrilateral elements. The optimal design was then extruded to achieve a cross-section with an aspect ratio of 1. Both models assume geometrically linear, small deformations. The material model is based on the small-strain shape-memory polymer constitutive model proposed by Baghani *et al.* [22]. To solve the numerical optimization problem, we use the gradient-based method of moving asymptotes (MMA) [23]. The design sensitivities required by the MMA algorithm are computed analytically using a transient adjoint formulation. Further details of the FEA model and the sensitivity analysis procedure can be found in [16].

There are two classes of kinematic elements that must be designed via topology optimization: the bending element and the twisting element. Once we have the material distribution for each class of element, we can simply rotate or flip the design to obtain angular actuations for all three rotational degrees of freedom, as shown in Figure 4. Note that during the final synthesis phase of the design framework, the algorithm will select from among six actuators at each point along the chain. However, all four bending actuators will have the same material distribution, and both twisting actuators will have the same material distribution. In addition to these six options, the optimizer may also insert a neutral element, whose shape remains unchanged during the activation process. This element can be used as an extender to shift the mechanism's kinematics along a given axis, and it is included in Figure 4(a) to illustrate the initial shape of all actuators prior to actuation.

The topology optimization procedure is used to obtain a precise distribution of the active and passive materials that will produce the desired rotational motion at the end of the thermomechanical programming cycle. For this task, the authors have developed a novel material representation scheme, which is an extension of the SIMP method, which is commonly used for elasticity problems [24]. In our case, we must continuously interpolate the elastic modulus, $E$, but also the viscosity coefficients, $\eta_r$ and $\eta_g$ (one for the glassy and rubbery phases of each design material), and the thermal expansion coefficients, $\alpha$. Hence, within each element, the effective value of a given material property, $\Psi$, is computed as an interpolation between the actual properties of the two active and passive design materials, $\Psi_1$ and $\Psi_2$ respectively. The interpolation formula is given in Eqn. (1).

$$\Psi_{\text{eff}} = \Psi_1 + \rho^p (\Psi_2 - \Psi_1). \tag{1}$$

Here, $\rho$ is the design variable assigned to the element in the FEA model. When $\rho = 0$, the element will have the properties of material 1 (i.e. $\Psi_{\text{eff}} = \Psi_1$), and when $\rho = 1$, the element will have the properties of material 2. The parameter $p$ is a constant that is used to *penalize* hybrid material states. When $p = 3$, this pushes the optimization search toward binary designs. Under our interpolated scheme, penalization is applied only when interpolating the elastic moduli, $E$. When interpolating all other material properties, the penalization is turned off by setting $p = 1$.

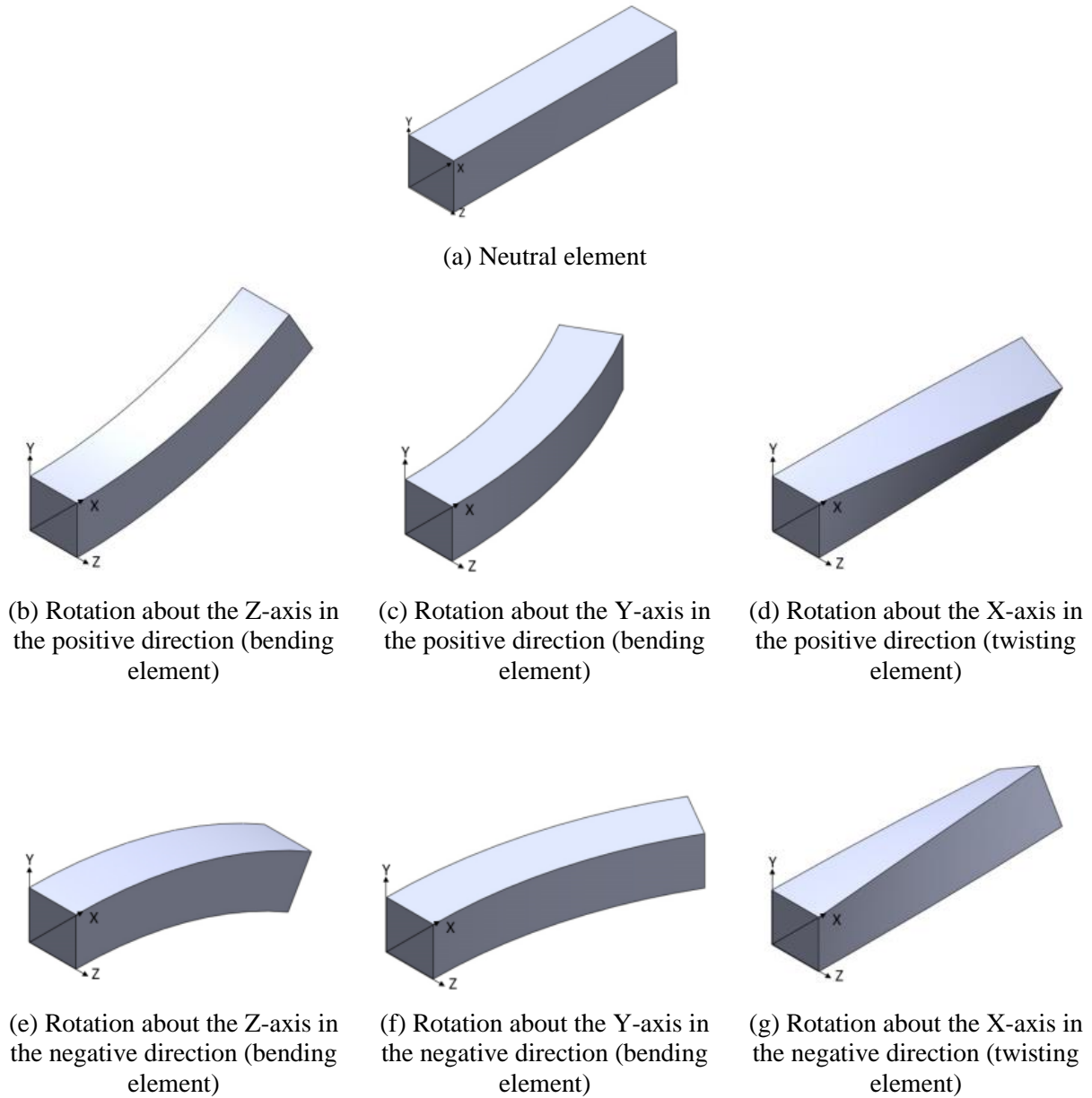

(a) Neutral element

(b) Rotation about the Z-axis in the positive direction (bending element)

(c) Rotation about the Y-axis in the positive direction (bending element)

(d) Rotation about the X-axis in the positive direction (twisting element)

(e) Rotation about the Z-axis in the negative direction (bending element)

(f) Rotation about the Y-axis in the negative direction (bending element)

(g) Rotation about the X-axis in the negative direction (twisting element)

**Figure 4:** Actuator options for the kinematic elements, which form the building blocks of the self-tying knot mechanism.

For the bending element, all motion occurs within $xy$-plane. Therefore, we can treat the material distribution problem as a two-dimensional problem and extrude the solution along the $z$-axis to obtain the full three-dimensional material distribution. Figure 5 shows the geometry and boundary conditions for the two-dimensional design domain used for the bending elements. This figure also shows the interpolated and extruded material distribution, which form the basis of the 3D CAD model that was ultimately fabricated using 3D printing. During the interpolation process, we begin with the material distribution field, represented by the variable $\rho$, which is piecewise constant within each finite element. This data is then interpolated to extract the precise location of the material interface to obtain a high-resolution representation of the internal material distribution. In this context, the material interface is selected as the contour or *level set* corresponding to $\rho = 0.5$.

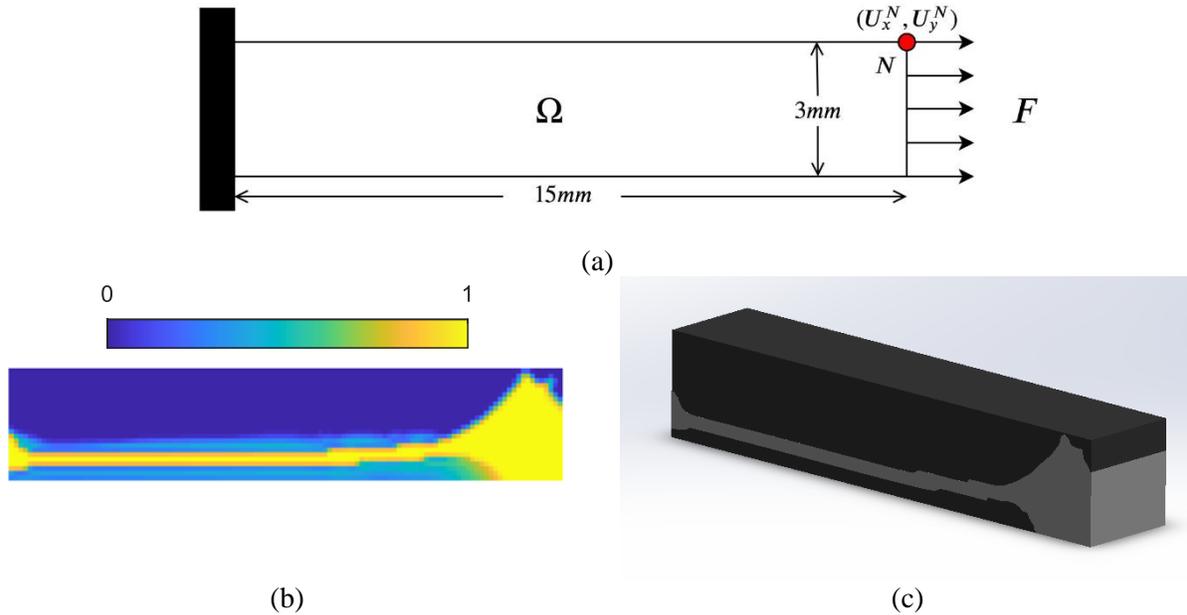

(a)

(b)            (c)

**Figure 5:** Topology optimization of the bending element: (a) Geometry and boundary conditions of the design domain. The design materials must be distributed within this rectangular domain. Note also that the axial force, $F$, is applied only during the initial stage of the thermomechanical programming cycle. (b) Optimized material distribution in pixel form. The blue region represents the active material and the yellow region represents the passive material. The color bar indicates the value of the design variable $\rho$. Each pixel in the image represents a single element in the finite element mesh. (c) The interpolated 3D material distribution after extrusion in the z-axis. Note that the interpolation process removes all intermediate density material appearing along the interface.

The optimization problem statement for the bending element is given in Eqn. (2). Note that we seek a material distribution, represented by the design variable, $\rho$, that will maximize the vertical displacement $U_y^N$ at the designated node, $N$ (marked with a red dot in Fig. 5, at the end of the thermomechanical cycle ($t = t^*$), while constraining the optimizer to ensure that the total volume of active material $V_{SMP1}$ does not exceed 70% of the total volume of the element. We impose a limit on the volume of the active material because this material is less stiff than the passive material. Therefore, it is necessary to limit the amount of active material available to the optimizer in order to ensure that all kinematic units have the necessary axial stiffness and that all portions of the mechanism are stretched by a similar amount during the thermomechanical programming cycle. The specific choice of a 70% volume fraction was chosen as a tradeoff between achieving a stiff design and a design that could produce large displacements, since the active material is what drives the morphing behavior.

$$\begin{aligned}&\underset{\rho}{\text{maximize}} && U_y^N|_{t=t^*}\\&\text{subject to} && V_{SMP^1}\leq 0.7,\qquad 0\leq\rho\leq 1\end{aligned}\qquad(2)$$

Figure 6 shows the geometry and boundary conditions for the design of the twisting element. This figure also contains the optimized material distribution generated by the algorithm, along with the interpolated material distribution within the 3D CAD model. Note that due to the high computational cost of the three-dimensional transient thermoelastic finite element analysis and the accompanying sensitivity analysis, we use a relatively coarse mesh for the initial topology optimization, however after the interpolation process, we obtain a sufficiently smooth material interface.

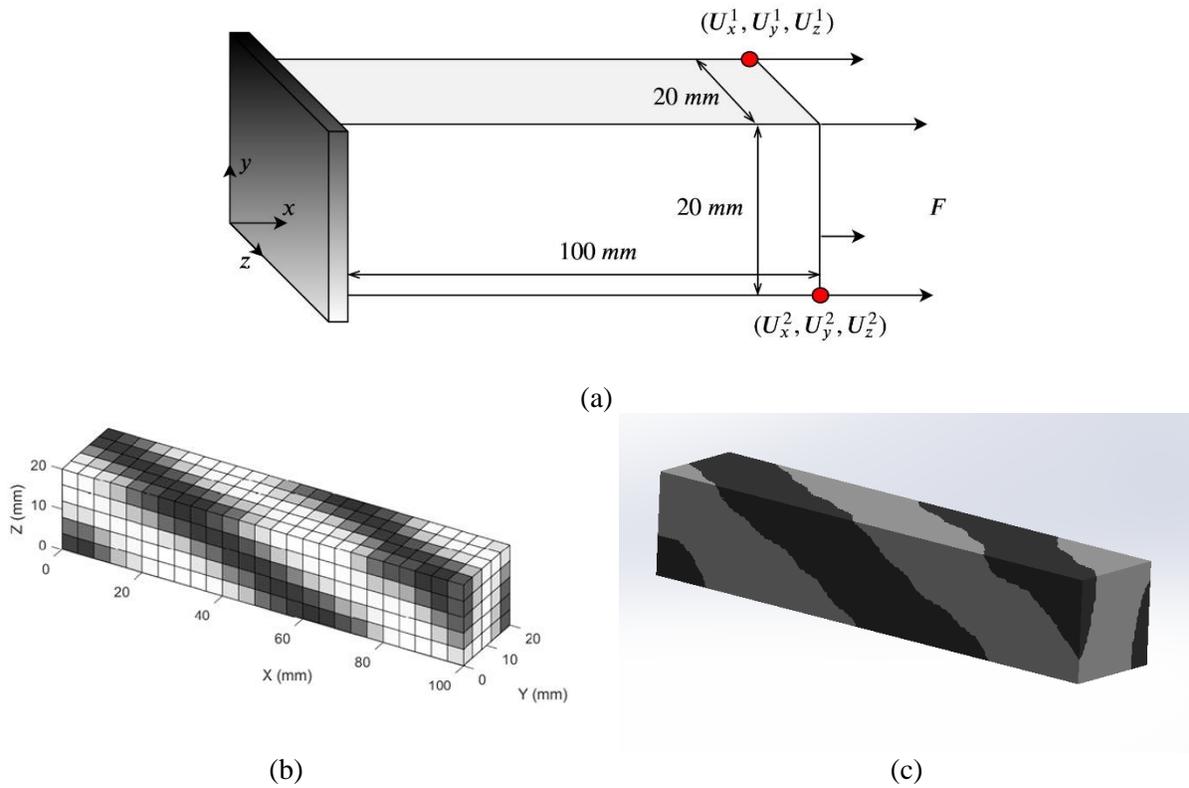

(a)

(b)                                                      (c)

**Figure 6:** Topology optimization of the twisting element: (a) Geometry and boundary conditions of the design domain. (b) Optimized material distribution in pixel form. The black region represents the active material and the white region represents the passive material. (c) The interpolated 3D material distribution.

Equation (3) shows the optimization problem statement for the design of the twisting element. Note that we seek to minimize the deflection of node 2 (the lower red dot in Fig. 6(a)) in the $z$ direction, thereby pushing this node inward. At the same time we constrain the $z$ deflection of node 1 to be a positive number. These two displacements combine to create a twisting motion at the free face of the element.

$$\begin{aligned}
&\underset{\rho}{\text{minimize}} && U_z^2|_{t=t^*} \\
&\text{subject to} && V_{SMP^1} \leq 0.7, \qquad 0 \leq \rho \leq 1 \\
& && U_z^1|_{t=t^*} > 0.8(U_z^1)_{initial}
\end{aligned} \qquad (3)$$

The two-dimensional topology optimization of the bending element as well as the 3D topology optimization of the twisting element is implemented in C++, with the PETSc library used for parallelization. Both sets of code are published in a public repository.

*Experimental Measurement and Characterization*

The bending and twisting structures are fabricated using the Stratasys Objet260 Connex 3D printer. The printer produces a wide range of *digital materials* by mixing two base materials: a stiff polymer *VeroWhite Plus* and a soft rubber-like elastomer *TangoBlack Plus* [7]. All the digital materials exhibit shape-memory behavior and each of the digital materials has a unique glass-transition temperature ($T_g$). We printed a subset of the available digital materials and experimentally determined their glass-transition temperatures. Finally, two digital materials, RGD8530 and FLX9895, were selected as the design materials for the bending and twisting structures. These materials were chosen because they have clearly distinct glass transition temperatures, allowing us to activate targeted sections of the mechanism using a uniform temperature field.

The bending and twisting structures were built with dimensions of $1 \times 0.2 \times 0.2$ cm so that their aspect ratios of 5:1:1 matched the designs obtained computationally. The structures with these particular dimensions were fabricated and subjected to the thermomechanical programming cycle, as shown in Figure 7, to determine their displacement response. The equipment used to characterize the displacement kinematics is described below.

1. **Temperature-controlled water bath**: This is used to apply a uniform temperature field to the shape-memory polymer samples over an extended period of time.
2. **Tensile device**: A global uniaxial strain is applied to the SMP samples during the deformation step of the thermomechanical programming cycle using an adjustable tensile device. The tensile device connects to two loops attached to either end of the morphing mechanism. The loops are made of VeroWhite, a stiff polymer with a high transition temperature. This material choice ensures that the connecting loop remains relatively rigid during the stretching of the mechanism.
3. **Water tank and water circulator with temperature control**: The stretched SMP structures are placed inside a temperature-regulated water tank and maintained at a fixed temperature using a water circulator during the reheating step for their motion activation.

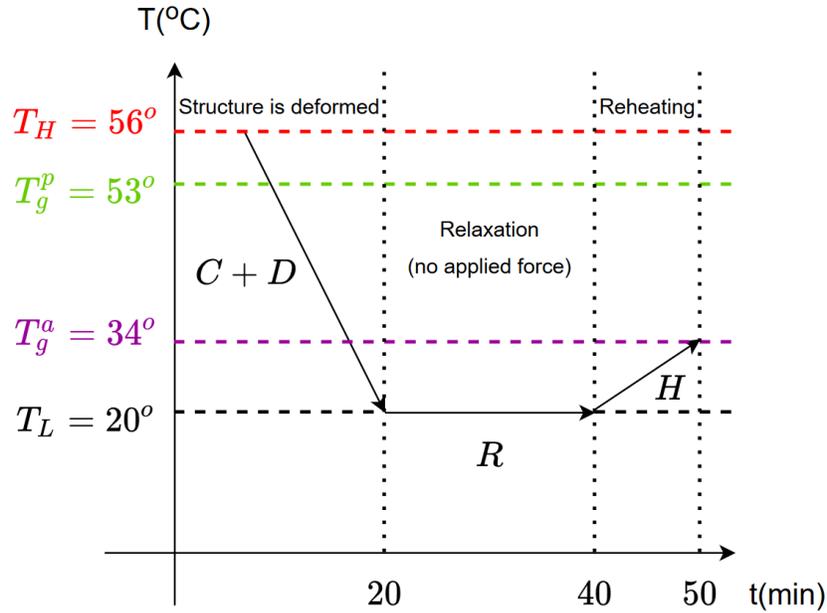

**Figure 7:** The thermomechanical programming cycle used for the experimental case studies. $T_H$ and $T_L$ represent the highest and lowest temperatures of the cycle. The values $T_g^p$ and $T_g^a$ represent the glass-transition temperatures of the active and passive SMP materials, respectively.

The thermomechanical programming cycle consists of three steps: cooling with deformation ($C + D$) of the specimen, relaxation ($R$), and reheating ($H$) of the specimen to obtain the desired shape change. The procedure is illustrated graphically in Fig. 7. Prior to starting the cycle, we heat the mechanism up to a temperature $T_H$, that is above the glass transition temperatures of both the active and passive SMP materials so that the entire specimen is in the rubbery phase. We then begin the programming cycle by applying a tensile load to stretch the specimen while cooling it to a temperature $T_L$ that is below the glass transition temperatures of both the active and passive materials. This cooling and stretching step lasts 20 minutes, during which both materials transition to their glassy phase. Next, during the relaxation step we release the applied forces while maintaining the specimen at its low temperate $T_L$ for an additional 20 minutes. During this step, the internal stresses in the material reduce the zero, however the material retains its stretched shape due to the strain fixity property of the shape memory polymers. In the third and final stage of the cycle, we raise the temperature to 34 degrees Celsius, which is the glass transition temperature of the active material. In this way, only the active material transitions back to the rubbery phase and therefore seeks to return to its original undeformed shape. The resulting disparity in strain between the two material regions causes the mechanism to exhibit the targeted shape change.

To measure the deflection angle of the bending element, a chain of 10 bending elements was 3D printed and the combined structure was subjected to the full thermomechanical programming cycle. The bending angle for a single element was determined by evaluating the bending angle of the combined structure and dividing by the number of elements. The bending angle was obtained by analysis of the digital image of the activated structure shown in Fig. 8(b). By taking this averaged measurement, we sought to reduce the impact of outliers and manufacturing defects. Similarly, the rotation angles for individual torsional specimens were calculated by subjecting a sequence of three torsional elements to the full thermomechanical programming cycle and calculating the average rotation angle.

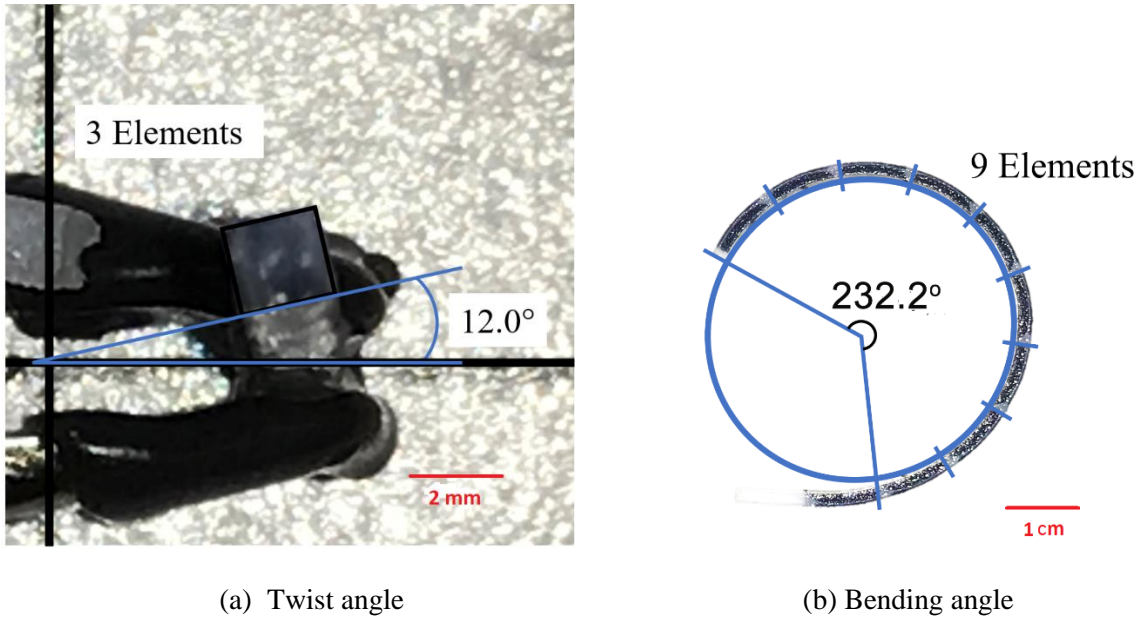

(a) Twist angle  (b) Bending angle

**Figure 8:** Experimental determination of the twist and bending angles of the kinematic elements.

For all kinematic elements, the rotation angle of the activated element is proportional to the strain applied during the thermomechanical programming cycle. For the twisting elements, the test samples were stretched to a strain of 13%, which produced a twist angle of 4° per element as shown in Fig. 8a. For the bending elements, eight different samples were tested with strains ranging between 9% and 16%. These results were then interpolated to obtain the expected bending angle for a strain of 13%, which was the applied strain of the self-tying knot during thermomechanical programming. This interpolation produced a bending angle of 25° for each individual bending element. These results (4° and 25°) were then used as the respective activation angles for the twisting and bending elements in the forward kinematic simulation described in the following section. Note that the finite element model predicted angular displacements of 3.5° and 33° for the torsional and bending elements respectively. This corresponds to an error of 12.5% for the torsional element and 32.0% for the bending element. The large discrepancy observed in the displacement for the bending beam is primarily due to the assumption of small strains in the finite element model. However, this error is mitigated by the fact that only the experimentally obtained displacements are used in the final kinematics model.

*Mechanism Synthesis*

The displacement and orientation of the tip of the knot is evaluated computationally using a forward kinematics model. The rotation angle of the bending and twisting elements used in the model are based on the experimentally measured rotation angles discussed above. An *ideal knot* shape is used as a target design to guide the algorithm. The path of the ideal knot is given by the parametric equation shown below.

$$x = 11.47(\sin(t) + 2\sin(2t))$$
$$y = 11.47(\cos(t) - 2\cos(2t)) \tag{4}$$
$$z = 4.13(-\sin(3t))$$

Note that the parameter $t$ ranges from $-3\pi/4$ to $3\pi/4$. The values of the coefficients in Eqn. 4 were selected so that the target knot size would remain within the volume of the 3D printer's maximum build envelope. The forward kinematics model is combined with a genetic algorithm to arrange the twisting and bending elements in an optimal sequence so that the end shape of the *trial knot* resembles the shape of the *ideal knot* as closely as possible. Note that because the knot is symmetric, we design only the half-knot, and reflect this solution about the root (base) of the knot to obtain the full knot path. The objective function for the optimizer consists of two terms as shown below:

$$y = C_0 P_{error} + C_1 Q_{error} \tag{5}$$

where $P_{error}$ represents the error between the Cartesian coordinates $\{x, y, z\}$ of the trial knot and the ideal knot at designated anchor points. In this study, the anchor points are chosen as the midpoint of the half-knot and the tip of the half-knot. The second term in Eqn. 5, $Q_{error}$, represents the error in the orientation of the two knots at the tip.

$$P_{error} = \sqrt{t_1 + t_2}$$
$$t_1 = (x_{t0} - x_t)^2 + (y_{t0} - y_t)^2 + (z_{t0} - z_t)^2$$
$$t_2 = w_m((x_{m0} - x_m)^2 + (y_{m0} - y_m)^2 + (z_{m0} - z_m)^2) \tag{6}$$
$$Q_{error} = \sqrt{(\phi_0 - \phi)^2 - (\theta_0 - \theta)^2}$$

The coefficients were set to $C_0 = 1.0$ and $C_1 = 5.0$ in order to appropriately scale the two error terms. The subscripts $m$ and $t$ refer to the midpoint and the tip of the knot respectively. The subscript 0 refers to the coordinates of the ideal knot and the constant coefficient was set to $w_m = 5.0$ to assign a greater weight to the midpoint error, since errors at the midpoint will magnify errors at the tip of the knot. The angles $\phi$ and $\theta$ capture the orientation of the knot segment expressed in spherical coordinates. Note that for the 13-segment half-knot (12 elements plus one neutral element), the *midpoint* of the ideal knot was approximated using the $\{x, y, z\}$ coordinates evaluated at $t = (3\pi/4)/(12/2 + 1)$. The forward kinematics algorithm loops over all the elements of the trial knot, evaluating the rotation matrices along each axis ($\boldsymbol{R}_x$, $\boldsymbol{R}_y$ and $\boldsymbol{R}_z$) within a local reference frame as shown below:

$$\boldsymbol{R}_x = \begin{pmatrix} 1 & 0 & 0 \\ 0 & \cos\alpha & -\sin\alpha \\ 0 & \sin\alpha & \cos\alpha \end{pmatrix} \tag{7}$$

$$\boldsymbol{R}_y = \begin{pmatrix} \cos\beta & 0 & \sin\beta \\ 0 & 1 & 0 \\ -\sin\beta & 0 & \cos\beta \end{pmatrix} \tag{8}$$

$$R_z = \begin{pmatrix} \cos\gamma & -\sin\gamma & 0 \\ \sin\gamma & \cos\gamma & 0 \\ 0 & 0 & 1 \end{pmatrix} \tag{9}$$

where $\alpha$, $\beta$ and $\gamma$ are rotational angles undergone by a 3D body about the $x$-axis, $y$-axis and $z$-axis respectively. The total rotation ($R_e$) of the 3D element is computed from the rotation along each of the axes. This is then used to evaluate the global rotation of the full sequence ($R$).

$$R_e = R_x R_y R_z = \begin{pmatrix} 1 & 0 & 0 \\ 0 & \cos\alpha & -\sin\alpha \\ 0 & \sin\alpha & \cos\alpha \end{pmatrix} \begin{pmatrix} \cos\beta & 0 & \sin\beta \\ 0 & 1 & 0 \\ -\sin\beta & 0 & \cos\beta \end{pmatrix} \begin{pmatrix} \cos\gamma & -\sin\gamma & 0 \\ \sin\gamma & \cos\gamma & 0 \\ 0 & 0 & 1 \end{pmatrix} \tag{10}$$

For each element in the sequence, offset distances along each axis ($\delta_x$, $\delta_y$ and $\delta_z$) are calculated for the displaced kinematic element.

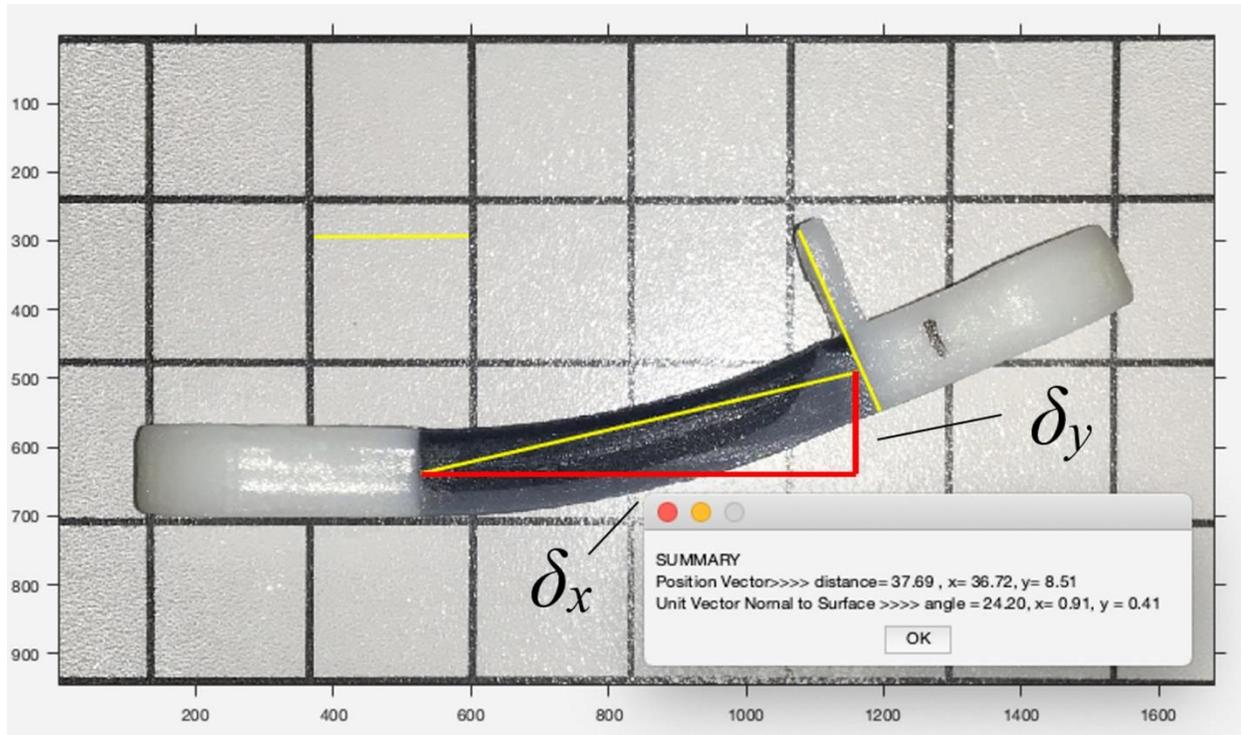

**Figure 9:** Offset distances shown on a single activated kinematic bending element.

For a bending element, shown in Fig. 9, the offset distances are given below.

$$\begin{aligned} \delta_x &= 0.98L \\ \delta_y &= 0.22L \\ \delta_z &= 0 \end{aligned} \tag{11}$$

Here, $L$ is the length of the element in the $x$-dimension prior to thermoelastic deformation. The offset distances for the twisting element are given below.

$$\delta_x = L$$
$$\delta_y = 0 \qquad (12)$$
$$\delta_z = 0$$

The offset distances are used to evaluate the global change in position vector ($\Delta X$) which is then used to calculate the position of the midpoint and tip of the full trial knot. The steps followed in the forward kinematics framework are shown in Algorithm 1 below.

**Algorithm 1:** Pseudocode for the forward kinematics analysis

```
R = eye(3,3) /* initialize rotation matrix as the identity (i.e. no
initial rotation)                                                  */
X = zeros(3,1) /* initialize the position vector                   */
for i ← 1,2,…,N do
   /*  loop over all the elements of the sequence                  */
   /*  Evaluate R_x, R_y and R_z given by Equation 7, 8 and 9      */
   /*  Form a vector with offset distances δ_x, δ_y and δ_z using
       Equation 11 and 12                                          */
   ΔX = R (δ_x, δ_y, δ_z)^T /* Evaluate change in position vector ΔX */
   R_e = R_z R_y R_x /* Evaluate the total rotation based on the current
   element using Equation 10                                       */
   R ← R R_e /* Update the total rotation matrix R                 */
   X ← X + ΔX /* Update the position vector X                      */
```

The forward kinematics algorithm is combined with MATLAB's genetic algorithm function "ga". The forward kinematics module evaluates the objective function for each trial knot design, which is then fed into the optimizer. The optimization algorithm was initially allotted 10 elements with which to construct the self-tying knot. If the optimization failed to generate a complete knot (where the tips penetrate the opposing loops on either side), then the number of elements was increased by one. In addition to requiring a complete knot, the algorithm also rejected any designs that caused *collisions* in which two segments of the knot would come into contact with one another at any point during the morphing process. To detect potential collisions the morphing motion was decomposed into 25 increments, and the Euclidean distances between all node pairings were evaluated at each increment.

**Results & Discussion**

The minimum number of elements with which the algorithm successfully produced a complete self-tying knot design with no collisions was 13. The optimized sequence for the half-knot is given in Fig. 10. In the figure, the identifiers $a - g$ correspond to the letters of the seven kinematic element options shown in Fig. 4.

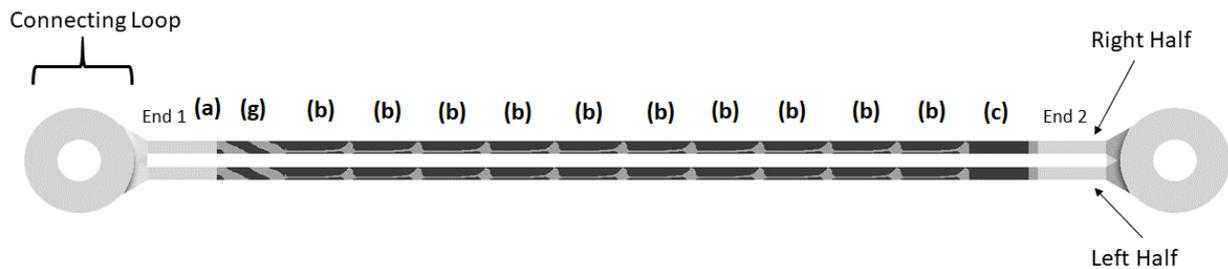

**Figure 10:** Offset distances shown on a single activated kinematic bending element.

As indicated in Fig. 10, the two halves of the designed knot are 3D printed together and they share a single connecting loop on both ends, for simultaneous thermomechanical programming. After stretching and relaxation, the connecting loops are removed with scissors and the two halves of the knot are mounted on a specialized stand prior to being placed in the heated water tank for final activation. The stand contains a two-sided flexible elastomer socket, and the two half-knots are slid into either side of the of the socket at *End 1* as indicated in Fig. 11.

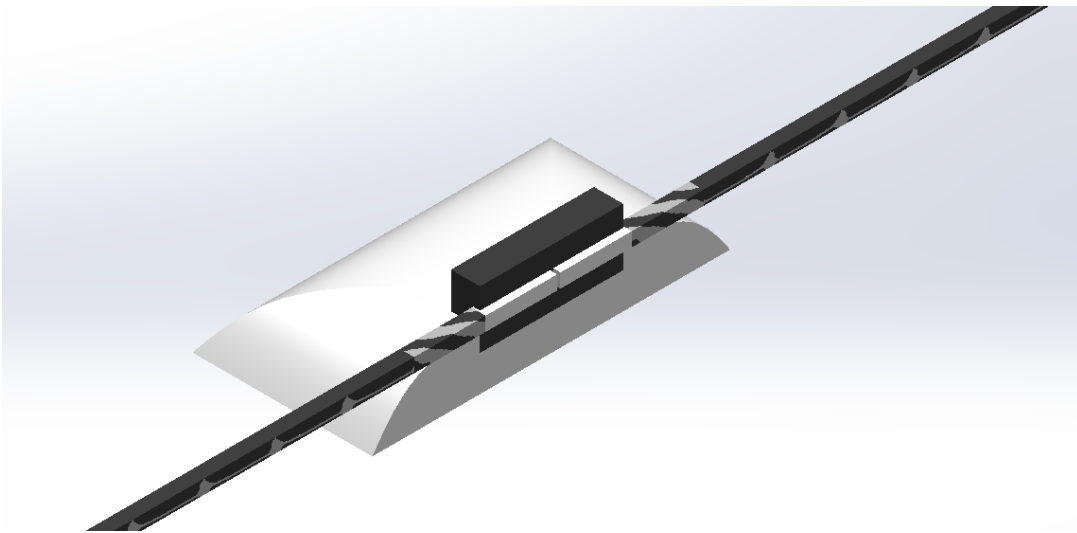

**Figure 11:** CAD model of the mounting apparatus used to hold the mechanism in place during activation (note that this image shows only the back half of the stand so that its cross-section is visible).

The activated self-tying knot shape is compared with the ideal knot shown in Fig. 12. The *blue* design represents the ideal knot generated mathematically, superimposed on the computationally generated self-tying knot shape, shown in *green*.

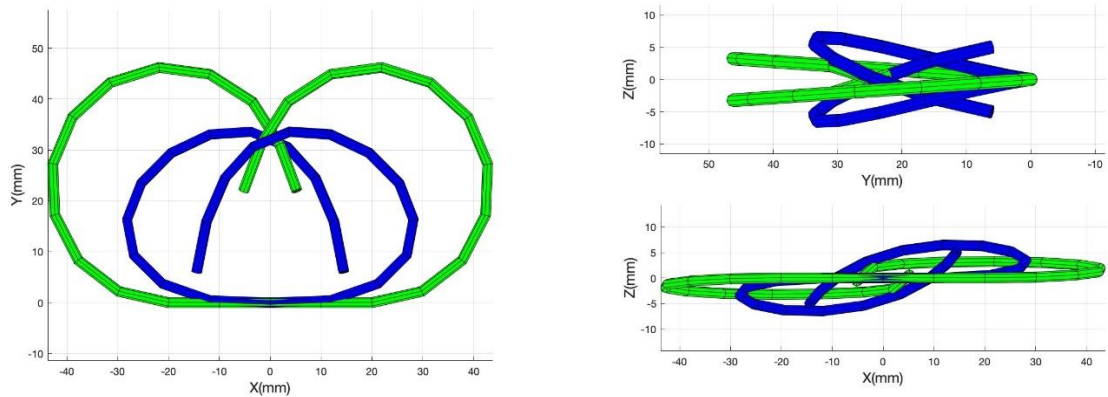

**Figure 12:** Comparison of the converged self-tying knot shape (green) with the mathematically generated ideal knot shape (blue).

Note that the geometry shown in Fig. 12 does not account for the effects of gravity. While the mechanism remains submerged in the water bath, the gravitational force is counteracted by the buoyant forces applied by the water and therefore we observe minimal sagging of the knot mechanism. Consequently, the photos of the submerged 3D printed knot (Fig. 14(b)) show good agreement with the forward kinematic simulation represented by Fig. 12. However, once the knot is removed from the water, gravitational effects become more apparent and the knot sags, as shown in Fig. 14(a). Figure 13 shows a computational simulation of the deformation of the activated knot mechanism when subject to gravity with no counteracting buoyant force. The simulation was performed using the commercial software package ANSYS with 13 space frame elements (one for each element in the forward kinematics model). Each element was assumed to be linearly elastic with 12 degrees of freedom per element. The shape of the deflected knot mechanism as determined by the simulation shows good agreement with the experimental results shown in Fig. 14(a).

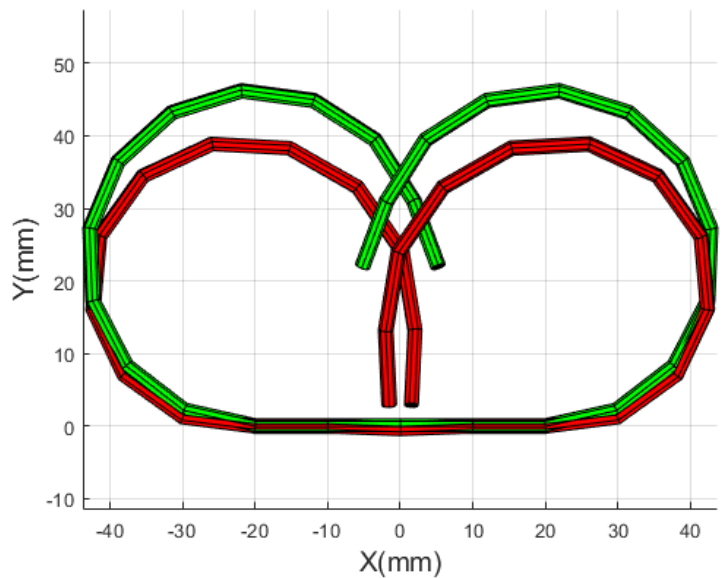

**Figure 13:** The activated self-tying knot with elastic deflection due to gravity (red) shown with the original undeflected knot (green).

Figure 14 shows the final shape of the actual 3D printed self-tying knot mechanism, along with a series of time-lapsed photos of the mechanism to illustrate the full range of motion.

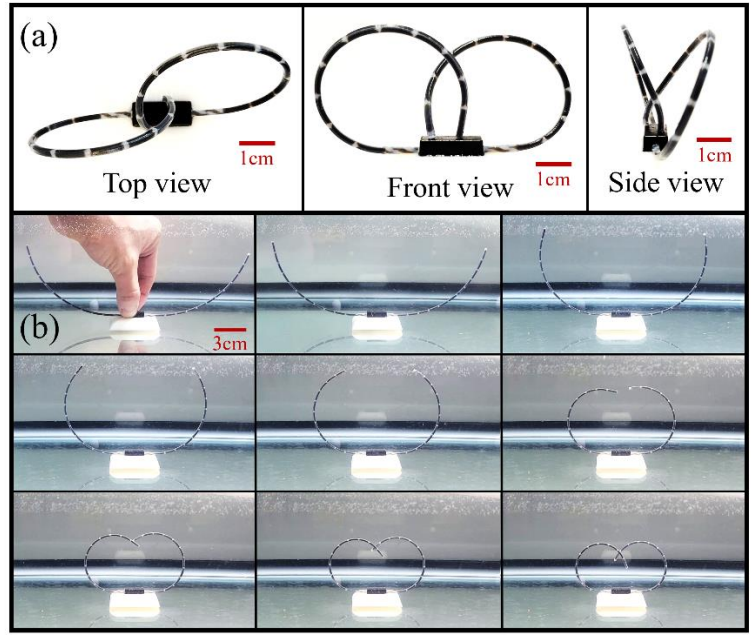

**Figure 14:** The final 3D printed self-tying knot mechanism; (a) The end shape of the self-tying knot shown from different viewing angles; (b) Time lapsed photos of the self-tying knot during the activation phase when the knot is submerged in a heated water bath. The images are separated by three-second intervals, with the full motion occurring over 24 seconds. Note that in (a) the mechanism has been removed from the water and is no longer subject to buoyancy forces, therefore its shape sags due to gravity.

**Conclusions**

The proposed design framework combines several novel features that enable unique functionality. The hierarchical design optimization approach allows for efficient exploration of the vast design space, which provides for enhanced programmability. The resulting mechanisms can therefore generate complex motion, all of which is systematically encoded into the material distribution. This capability is also enabled by our hybrid approach, in which experimental measurement is embedded into the computational design framework. Secondly, by replacing electromechanical components with material-based actuation, we make the mechanisms more lightweight, more robust to failure, and readily manufacturable via 3D printing. Furthermore, these material-based mechanisms are highly miniaturizable. This could facilitate the creation of micro- and nanorobots, which are particularly useful for in vivo biomedical applications such as targeted drug delivery [25]. Lastly, the lack of a central processor means that both sensing and actuation occur locally, allowing for rapid response to environmental changes.

In their 2012 book *Fabricated*, Kurman and Lipson [26] speculate that 3D printing will revolutionize the way we procure everyday household items such as toothbrushes, which can now be printed at home on demand. Advances in 4D printing may take this a step further by allowing consumers to 3D print previously unseen items such as self-tying shoelaces and combs with adaptive bristle density. Advances like the ones presented in this study could make possible an entirely new class of technologies that rely on programmable material-based robotic systems.

More importantly, this capability could provide a key ingredient necessary for realizing aspirational life-saving technologies such as artery-clearing microrobots, self-tying sutures, and many other disruptive technologies that today seem unimaginable.

**Data Availability**

All computer code used to generate the results of this study is publicly available via the following online repository: https://github.com/bhttchr6/STK_codes

[13] K. A. James, G. J. Kennedy and J. R. Martins, "Concurrent aerostructural topology optimization of a wing box," *Computers and Structures,* vol. 134, pp. 1-17, 2014.

[14] J. V. Carstensen, R. Lotfi, J. K. Guest, W. Chen and J. Schroers, "Topology Optimization of Cellular Materials With Maximized Energy Absorption," in *ASME 2015 International Design Engineering Technical Conferences*, Boston, Massachusetts, 2015.

[15] K. Maute, A. Tkachuk, J. Wu, H. J. Qi, Z. Ding and M. L. Dunn, "Level Set Topology Optimization of Printed Active Composites," *J. Mech. Des.,* vol. 137, no. 11, p. 111402 (13 pages), 2015.

[16] A. Bhattacharyya and K. James, "Topology optimization of shape-memory polymer structures with programmable morphology," *Structural and Multidisciplinary Optimization,* vol. 63, p. 1863–1887, 2021.

[17] G. Sossou, F. Demoly, H. Belkebir, H. J. Qi, S. Gomes and G. Montavon, "Design for 4D printing: Modeling and computation of smart materials distributions," *Materials & Design,* vol. 181, p. 108074, 2019.

[18] X. Sun, L. Yue, L. Yu, H. Shao, X. Peng, K. Zhou, F. Demoly, R. Zhao and H. J. Qi, "Machine Learning-Evolutionary Algorithm Enabled Design for 4D-Printed Active Composite Structures," *Advanced Functional Materials,* vol. 32, no. 10, p. 2109805, 2021.

[19] Z. Kang and K. James, "Multimaterial Topology Optimization for Elastic and Thermal Response," *International Journal for Numerical Methods in Engineering, 117(10):1019-1037, 2019.,* vol. 117, no. 10, pp. 1019-1037, 2019.

[20] M. A. Haque, N. V. Lavrik, D. Hensley, D. P. Briggs and N. McFarlane, "Carbonized Polymer for Joule Heating Processing Towards Biosensor Development," in *2021 43rd Annual International Conference of the IEEE Engineering in Medicine & Biology Society (EMBC)*, Mexico, 2021.

[21] R. P. Paul, Robot Manipulators: Mathematics, Programming, and Control : the Computer Control of Robot Manipulators, Cambridge, MA: MIT Press, 1981.

[22] M. Baghani, R. Naghdabadi, J. Arghavani and S. Sohrabpour, "A thermodynamically-consistent 3D constitutive model for shape memory polymers," *International Journal of Plasticity,* vol. 35, pp. 13-30, 2012.

[23] K. Svanberg, "The method of moving asymptotes—a new method for structural optimization," *International Journal for Numerical Methods in Engineering,* vol. 24, no. 2, pp. 359-373, 1987.

[24] M. P. Bendsøe and O. Sigmund, "Material interpolation schemes in topology optimization," *Archive of Applied Mechanics,* vol. 69, pp. 635-654, 1999.

[25] M. Hu, X. Ge, X. Chen, W. Mao, X. Qian and W.-E. Yuan, "Micro/Nanorobot: A Promising Targeted Drug Delivery System," *Pharmaceutics,* vol. 12, no. 7, p. 665, 2020.

**Acknowledgements**

This work was funded by the National Science Foundation through grant # 1663566.

**Author Contributions**

A.B. contributed to conception of the original idea, method development and implementation, and manuscript preparation. J.-Y. K. contributed to method development and implementation, conducting laboratory experiments, and manuscript editing. L.R.A. contributed to method development and implementation, and manuscript editing. K.A.J. contributed to conception of the original idea, method development and implementation, and manuscript preparation.

**Materials and Correspondence**

Correspondence and inquiries should be addressed to Kai A. James (kai.james@gatech.edu).